\documentclass[twocolumn,iop,preprint]{emulateapj}

\usepackage[mathlines,displaymath]{lineno}
\usepackage{graphicx}
\usepackage{multirow}
\usepackage{lineno}
\usepackage{subfigure}
\usepackage{color}
\usepackage{nicefrac}
\usepackage{color}
\usepackage{amssymb}
\usepackage{amsmath}
\usepackage{enumerate}
\usepackage{hyperref}
\usepackage{xspace}

\newcommand{\Mesz}{{M\'esz\'aros}}
\newcommand{\XRT}{\textnormal{\tiny \textsc{XRT}}}
\newcommand{\LAT}{\textnormal{\tiny \textsc{LAT}}}
\newcommand{\SSC}{\textnormal{\tiny \textsc{SSC}}}
\newcommand{\KN}{\textnormal{\tiny \textsc{KN}}}
\newcommand{\synch}{\textnormal{\tiny \textsc{synch}}}
\newcommand{\BB}{\textnormal{\tiny \textsc{BB}}}

\newcommand{\Fermi}{\emph{Fermi}\xspace}
\newcommand{\fermi}{\emph{Fermi}\xspace}

\def\fermi{{\it Fermi}}

\def\vFv{\nu F_{\nu}}
\def\Ep{E_{\rm p}}
\def\G0{\Gamma_{0}}
\def\q3{q_{-3}}
\def\Eiso{9.1\cdot10^{53}}
\def\xd{3.4\cdot10^{16}}

\def\p0{$\pi^{\rm 0}$}

\def\de{$^{\circ}$\xspace}

\def\p8{\texttt{Pass8}}
\def\p7{\texttt{Pass7REP}}

\newcommand*\patchAmsMathEnvironmentForLineno[1]{%
  \expandafter\let\csname old#1\expandafter\endcsname\csname #1\endcsname
  \expandafter\let\csname oldend#1\expandafter\endcsname\csname end#1\endcsname
  \renewenvironment{#1}%
     {\linenomath\csname old#1\endcsname}%
     {\csname oldend#1\endcsname\endlinenomath}}% 
\newcommand*\patchBothAmsMathEnvironmentsForLineno[1]{%
  \patchAmsMathEnvironmentForLineno{#1}%
  \patchAmsMathEnvironmentForLineno{#1*}}%
\AtBeginDocument{%
\patchBothAmsMathEnvironmentsForLineno{equation}%
\patchBothAmsMathEnvironmentsForLineno{align}%
\patchBothAmsMathEnvironmentsForLineno{flalign}%
\patchBothAmsMathEnvironmentsForLineno{alignat}%
\patchBothAmsMathEnvironmentsForLineno{gather}%
\patchBothAmsMathEnvironmentsForLineno{multline}%
}
\begin{document} 
%\linenumbers

\title{An External Shock Origin of GRB~\textit{141028A}}
%\title{The Prompt Emission of GRB~\textit{141028A}:An External Shock Origin?}
%\title{The Working Dead: An External Shock Origin for the Prompt Emission of GRB~\textit{141028A}}
\author{J.~Michael~Burgess\altaffilmark{1,2,$\dag$}, %
Damien~B\'egu\'e\altaffilmark{1,2,$\ddag$}, %
Felix~Ryde\altaffilmark{1,2}, %
Nicola~Omodei\altaffilmark{3}, %
Asaf~Pe'er\altaffilmark{4}, %
J.~L.~Racusin\altaffilmark{5}, %
A.~Cucchiara\altaffilmark{5} %
}
\altaffiltext{1}{The Oskar Klein Centre for Cosmoparticle Physics,
  AlbaNova, SE-106 91 Stockholm, Sweden}
\altaffiltext{2}{Department of Physics, KTH Royal Institute of Technology, AlbaNova University Center, SE-106 91 Stockholm, Sweden}
\altaffiltext{$\dag$}{jamesb@kth.se}
\altaffiltext{$\ddag$}{damienb@kth.se}
\altaffiltext{3}{W. W. Hansen Experimental Physics Laboratory, Kavli
  Institute for Particle Astrophysics and Cosmology, Department of
  Physics and SLAC National Accelerator Laboratory, Stanford
  University, Stanford, CA 94305, USA.}  
\altaffiltext{4}{Physics Department, University College Cork, Cork, Ireland}
\altaffiltext{5}{NASA Goddard
  Space Flight Center, Greenbelt, MD 20771, USA}

%\normalsize{$^\ast$To whom correspondence should be addressed. Email:}\\
%\normalsize{jamesb@kth.se (J.M.B;}\\

% Include the date command, but leave its argument blank.

%\date{}

%%%%%%%%%%%%%%%%% END OF PREAMBLE %%%%%%%%%%%%%%%%

%Double-space the manuscript.

\begin{abstract}
  The prompt emission of the long, smooth, and single-pulsed gamma-ray
  burst, GRB~\textit{141028A}, is analyzed under the guise of an
  external shock model. First, we fit the $\gamma$-ray spectrum with a
  two-component photon model, namely synchrotron+blackbody, and then
  fit the recovered evolution of the synchrotron $\vFv$ peak to an
  analytic model derived considering the emission of a relativistic
  blast-wave expanding into an external medium. The prediction of the
  model for the $\vFv$ peak evolution matches well with the
  observations. We observe the blast-wave transitioning into the
  deceleration phase. Further we assume the expansion of the
  blast-wave to be nearly adiabatic, motivated by the low magnetic
  field deduced from the observations. This allows us to recover
  within an order of magnitude the flux density at the $\vFv$ peak,
  which is remarkable considering the simplicity of the analytic
  model. Across all wavelengths, synchrotron emission from a single
  forward shock provides a sufficient solution for the observations.
  Under this scenario we argue that the distinction between
  \textit{prompt} and \textit{afterglow} emission is superfluous as
  both early and late time emission emanate from the same source.
  While the external shock model is clearly not a universal solution,
  this analysis opens the possibility that at least some fraction of
  GRBs can be explained with an external shock origin of their prompt
  phase.
\end{abstract}


\section{Introduction}
\label{sec:intro}
Identifying the origin of the dynamical evolution of gamma-ray burst
(GRB) outflows is an unsolved issue, critical to the understanding of
both the energetics and spectra of these events. One idea is that the
emission is the result of synchrotron radiation from an external
forward shock propagating into the external circumburst medium (CBM)
\citep{Cavallo:1978,Rees:1992,Meszaros:1993,Chiang:1999,Dermer:1999,Mitman:1999}. This
mechanism should produce smooth $\gamma$-ray pulses with durations on
the order of a few seconds for typical GRB parameters. However, the
short-time variability (on the order of a few milliseconds) of many
GRB light curves ruled out this postulate as a universal mechanism
\citep[e.g.][]{Sari:1997,Kobayashi:1997,Walker:2000} and gave favor to
several alternative hypotheses including models that consist of
rapid internal shocks in an unsteady outflow \citep{Rees:1994} or
magnetic reconnection
\citep[e.g.,][]{Spruit:2001,Drenkhahn:2002,Zhang:2011} to account for
the observed emission.

Still, there do exist long and temporally smooth GRBs with typical
variability time scales larger than a few seconds
\citep[see][]{Golkhou:2014,Golkhou:2015} that do not violate the
variability constraints of the external shock model and can be tested
via their spectral evolution as to whether they conform to the
well-established predictions made by this model. The simplicity of the
model affords it the ability to be tested both spectrally and
temporally, a feature unique to the external shock model. The dynamics
and spectra of the internal shock model have been simulated
\citep{Daigne:1998}, but the dynamics rely on assumed and degenerate
configurations from variations in the wind (e.g. the radial
distribution of the Lorentz factors), forbidding the formulation of
unique predictions that can be identified in the data. Therefore, it
is currently impossible to test the internal shock model in the manner
presented here without severe degeneracies. Additionally, the internal
shock model has trouble efficiently converting the internal kinetic
energy of a GRB into radiation, which is challenging when trying to
explain the extreme luminosities observed \citep{Kobayashi:1997}.

Herein, we analyze the bright, long, single-pulsed
GRB~\textit{141028A} and find several clues for an external
shock origin of its emission. We fit the GRB's time-resolved spectra
with a slow-cooled synchrotron+blackbody model \citep{Burgess:2014}
and examine the evolution of the spectra. The evolution of the
synchrotron $\vFv$ peak ($\Ep$) is fit with an analytic physical model
predicted by \citet{Dermer:1999}. From this fit, we obtain physical
parameters such as the coasting Lorentz factor and CBM radial profile
which can then be used to predict how the flux of the prompt emission
should evolve. Comparing these predictions to the data enables us to
test the validity of the model in several ways.

The article is organized in the following manner. In Section
\ref{sec:esm}, we introduce the formalism of the external shock model
to derive a function for $\Ep(t)$ to fit to data. In Section
\ref{sec:analysis}, the observations and spectral analysis are
introduced.  Section \ref{sec:esa} details our application of the
external shock model to the data. The parameters resulting from the
analysis are used to make further predictions regarding the flux
evolution of the prompt phase of the GRB. From there, we analyze the
photospheric emission. In Sections \ref{sec:afterglow} and
\ref{sec:lat}, we discuss the compatibility of our results with the
high and low-energy late-time observations.

\section{The External Shock Model}
\label{sec:esm}

The external shock model is built upon the blast-wave evolution
derived in \citet{Blandford:1976} which tracks the evolution of a
relativistically expanding fireball into an external medium.  The
equations can be applied to GRBs by assuming some fraction of the
electrons in the shocked external medium is accelerated to high
energies by the shock wave and radiates a fraction of the kinetic
energy away via synchrotron radiation
\citep{Cavallo:1978,Rees:1992}. In this work, we use the analytic
formalism developed in \citet{Dermer:1999} to fit the spectral
evolution of the emission. We briefly review the main equations
required and refer the reader to \citet{Dermer:1999,Chiang:1999} for
more details on the model.

The blast-wave is assumed to expand into an external circumburst
medium (CBM) with a radial density evolution modeled as a power law,

\begin{equation}
  \label{eq:cbm}
n(x) = n_0 x^{-\eta} \; {\rm cm^{-3}}  
\end{equation}

\noindent where $n_0$ is the initial density and $\eta$ describes the
radial morphology of the CBM such that $\eta=0$ is a constant density
and $\eta=2$ describes a stellar wind. The dimensionless radial coordinate is
$x=\nicefrac{r}{r_{\rm d}}$ where, following the convention that a quantity
$w=w_n10^n$,

\begin{equation}
  \label{eq:xd}
  r_{\rm d} = 5.4 \cdot 10^{16}\left[ \frac{(1-\nicefrac{\eta}{3})E_{0, 54}}{n_{0,2}\Gamma_{0,2}^2}\right]^{\nicefrac{1}{3}}\; {\rm cm}
\end{equation}

\noindent is the radius at which the blast has swept up a significant
amount of mass ($\propto\Gamma^{-1}$) to begin decelerating
\citep{Rees:1992}. Following the solution of \citet{Blandford:1976}
the evolution of the bulk Lorentz factor ($\Gamma$) of the blast-wave is
modeled as a broken power law consisting of a coasting phase followed
by a deceleration phase:

\begin{equation}
\label{eq:gamma}
\Gamma(x)=\left\{
\begin{array}{ll}
 \G0	& \;x<1 \\
 \G0 x^{-g}	& \;1 \le x
\end{array}
\right. 
\end{equation}

\noindent where $\G0$ is the coasting Lorentz factor and $g$ is the
radiative regime index. For a constant density ($\eta=0$) CBM,
$g=3,\nicefrac{3}{2}$ indicate the fully radiative and non-radiative
(adiabatic) expansion regimes respectively \citep{Blandford:1976}. A
fraction of the blast energy is dissipated in the shock and
accelerates electrons to high energies, which subsequently radiate
this away via synchrotron radiation. Following the parameterization of
\citet{Dermer:1999}, the temporal evolution of the $\vFv$ peak energy,
$\Ep$, of this synchrotron radiation can be modeled as

\begin{equation}
  \label{eq:ep}
\Ep(t) = \mathcal{E}_0 \left[ \frac{\Gamma(x)}{\G0}   \right]^4 x^{-\nicefrac{\eta}{2}} \; {\rm keV}  
\end{equation}
where 
\begin{equation}
  \label{eq:e0}
  \mathcal{E}_0 = 15\;  \frac{n_{0,2}^{\nicefrac{1}{2}} q_{-3}  \Gamma_{0,2}^4}{1+z} \; \text{keV}
\end{equation}
is the observed $\Ep$ at the observed deceleration time,
\begin{equation}
  \label{eq:td}
  t_{\rm d} = \frac{r_{\rm d}}{\G0^2 c}(1+z) \text{.}
\end{equation}

\noindent Here, $c$ is the speed of light and $q$ parameterizes the
magnetic field and shock acceleration microphysics such that $q\equiv
\left[ \epsilon_{\rm B}(\nicefrac{r_{\rm s}}{4})
\right]^{\nicefrac{1}{2}} \epsilon_{\rm e}^2$. Here $\epsilon_{\rm B}$
and $\epsilon_{\rm e}$ are the magnetic and electron equipartition
factors and $r_{\rm s}\simeq4$ is the shock compression ratio. For
values expected in GRBs and to coincide with our choice of
slow-cooling synchrotron to fit the spectra, if we take $\epsilon_{\rm
  B}\simeq 10^{-4}$ and $\epsilon_{\rm e} \lesssim 0.5$ then we have
values of $q \simeq 10^{-3}$. While these values of $\epsilon_{\rm B}$
are lower than the typically assumed values of $\sim 10^{-2}$, recent
studies \citep{Lemoine:2013,Santana:2014} find that the values can be
much lower and we therefore follow these works. For the analysis, we
will allow values $\epsilon_{\rm B}\lesssim 10^{-2}$ to cover the
range of typically assumed values.

Additionally, we can write the measured $\vFv$ peak luminosity as
\begin{equation}
  \label{eq:pt}
  P_{\rm p}(t)=\Pi_0\left\{
\begin{array}{ll}
 x^{2-\eta}	& \;\;0\le x<1 \\
 x^{2-\eta -4g}	& \;\;1\le x< \G0^{\nicefrac{1}{g}}
\end{array}
\right.
\end{equation}
where
\begin{equation}
  \label{eq:p0}
  \Pi_0 \propto \frac{(2g-3+\eta)m_{\rm p} c^3 \G0^4n_0 A_0}{2g(1+z)^2} \; {\rm erg\;\;s^{-1}.}
\end{equation}

\noindent Here, $m_{\rm p}$ is the proton mass and
$A_0$ is the blast-wave area factor. The proportionality in
Equation~\ref{eq:p0} comes from the fact that in \citet{Dermer:1999}
an empirical photon model is used to create light curves and here we
will be using synchrotron emission directly to fit the time-resolved
spectra of GRB~\textit{141028A} (see Section~\ref{sec:analysis}).

\begin{figure}[h]
  \centering
  \includegraphics{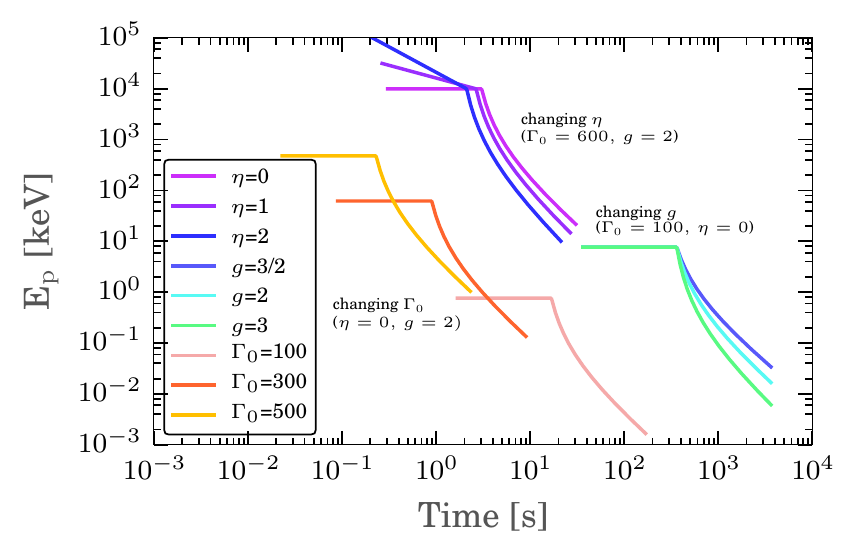}
  \caption{Demonstrating the effect of the physical parameters on the
    evolution of $\Ep$ with time. It is easy to see that $\G0$ has the
    strongest effect on the timescale of the burst, while $\eta$
    affects the early evolution of $\Ep$ and $g$ the late evolution.}
  \label{fig:demo}
\end{figure}

With these equations, the external shock model can predict several
observable quantities, namely, the evolution of $\Ep$ and $P_{\rm p}$
which we test below in Section \ref{sec:eppp}. The effects on the
$\Ep$ evolution of different parameters is shown in
Figure~\ref{fig:demo}. The parameters have the following dominant
effects:
\begin{itemize}
\item $\G0$: sets the break time
\item $\eta$: sets the slope of early $\Ep$ decay
\item $g$: sets the slope of late $\Ep$ decay.
\end{itemize}

\noindent These parameters, along with $q$ which acts as a
normalization, will be the free parameters that will be determined by
fitting the observed evolution of $\Ep$ to Equation~\ref{eq:ep}. In
reality, the range of values for $g$ are determined by the value of
$\eta$ \citep{Dermer:1999}. Since we cannot know a priori the value of
$\eta$, we treat $g$ as an independent parameter. It is important to
note that this analytic approximation to the blast-wave evolution
fails to capture some aspects that the full numerical solution
exhibits \citep{Chiang:1999}. Most notable is the smooth transition
from coasting to decelerating at the deceleration radius ($r_{\rm
  d}$). The formalism we adopt will serve as a proof of concept that
will be further applied to a larger sample and improved upon with a
full numerical treatment.

\section{Observations of GRB~\textit{141028A}}
\label{sec:analysis}
\subsection{Data Acquisition}
\begin{figure}[h]
  \centering

  \includegraphics{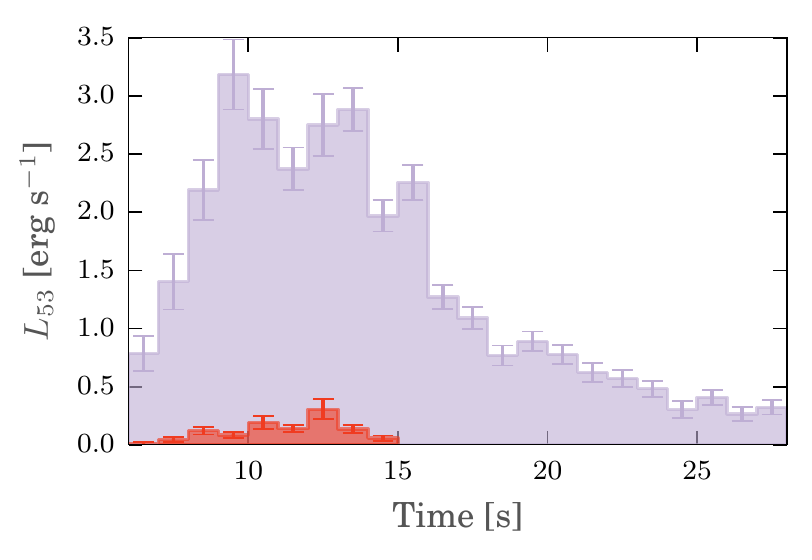}
  \caption{The luminosity lightcurve of GRB~\textit{141028A}
    consisting of the synchrotron ({\it purple}) and blackbody ({\it
      red}) components.}
  
\label{fig:flux}
\end{figure}

Rapid variability poses a problem for the external shock model;
therefore, bright, long, single-pulsed GRBs provide the most viable
candidates for being produced by external shocks. GRB~\textit{141028A}
is an example of this class of GRB (see
Figure~\ref{fig:flux}). GRB~\textit{141028A} was discovered by the
{\it Fermi} Gamma-ray Burst Monitor (GBM)
\citep{gcn:16971,Meegan:2009} and the bright GRB triggered an
autonomous repoint of the {\it Fermi} spacecraft to optimize the Large
Area Telescope (LAT) \citep{LATinstrument} for follow-up observations.
In ground analysis, LAT also localized and detected the GRB
\citep{gcn:16969} until $\sim 10^3$ s after the trigger. The
localization led to a target of opportunity observation by the
narrow-field {\it Swift} instruments (the X-ray Telescope, XRT; and
the Ultraviolet Optical Telescope, UVOT) \citep{Gehrels:2004}.  The
X-ray and optical afterglow was detected by XRT \citep{gcn:16978} and
UVOT \citep{gcn:16979}, and subsequently by many different
ground-based facilities, including a measurement of a redshift of
z=2.332 with the Very Large Telescope/X-Shooter instrument
\citep{gcn:16983}.  We collected optical/NIR photometry from GCN
circulars, and constructed a multi-band SED using GROND data
\citep{gcn:16977}, and light curves using r' and i' filters (GROND,
RATIR: \citet{gcn:16980}, P60: \citet{gcn:16989}, LCOGTN:
\citet{gcn:16985}). The XRT light curve and spectra were obtained from
the XRT Team Repository \citep{Evans:2007,Evans:2009}. We triggered a
pre-approved late-time Chandra target of opportunity observation to
constrain the properties of the break in the X-ray afterglow hinted at
by XRT.  The 40 ks Chandra observation was analyzed with CIAO v4.6,
yielding a faint detection which was converted to flux using the fit
to the XRT spectrum. The X-ray and optical afterglow observations
began at $\sim10^4$ s after the GBM trigger. It is therefore
impossible to observe the continuous evolution of the flux from the
prompt to afterglow phase at low-energy.

\subsection{Spectral and Temporal Analysis}
For the prompt emission, we use GBM time-tagged event (TTE) data and
$Fermi$ LAT low-energy (LLE) data to perform spectral analysis. The
LLE technique is an analysis method designed to study bright transient
phenomena in the 30 MeV - 1 GeV energy range, and was successfully
applied to \fermi-LAT GRBs \citep{2013ApJS..209...11A} and solar
flares \citep{2012ApJ...745..144A, 2014ApJ...789...20A}.  The idea
behind LLE is to maximize the effective area below $\sim$ 1 GeV by
relaxing the standard analysis requirement on background
rejection. The first five seconds after the trigger time are excluded
because the count rate was too low to constrain a spectral model.

\begin{figure*}[h]
  \centering
\subfigure[]{
  \includegraphics{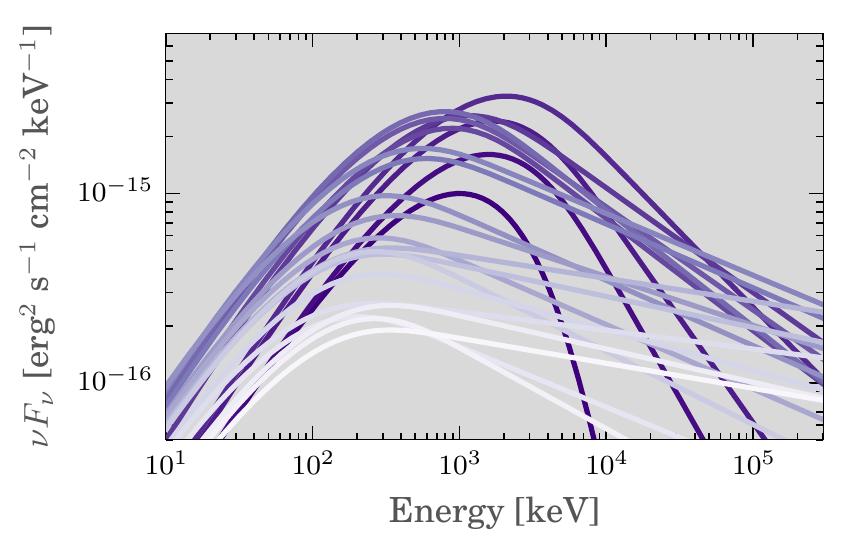}}\subfigure[]{ \includegraphics[scale=1]{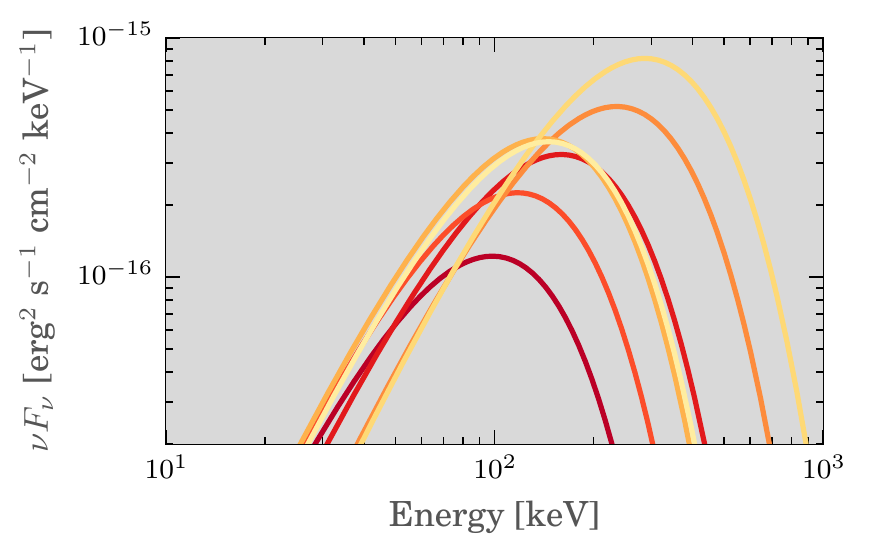}}
\caption{The time-resolved $vFv$ spectra in 1 s intervals of the synchrotron (5-27 s) (a) and
  blackbody (5-15 s) (b) components from the GBM+LLE fits. The synchrotron
  spectrum evolves in time from \textit{purple} to \textit{white}
  while the blackbody evolves from \textit{red} to \textit{yellow}.}
  \label{fig:spectrum}
\end{figure*}

Using the method of \citet{Gao:2012}, we calculated the variability
components of GRB~\textit{141028A} during the prompt emission. We find
the dominant component to be $t_{\rm var}=24.7/(1+z)$~s and an
insignificant fast component with $t_{\rm var}=3.8/(1+z)$~s. These
values place the burst safely within the range of what can be expected
by an external shock. We use time bins of 1 s to be sure to bin below
the fast variability component. To fit the time-resolved spectra, we
employ a two-component model consisting of synchrotron emission from
an incompletely cooled electron distribution (slow-cooled) and a
blackbody \citep{Burgess:2011,Burgess:2014} (see Figure
\ref{fig:spectrum}). In \citet{Burgess:2014}, it was shown that single
pulsed GRBs have spectra that are compatible with slow-cooled
synchrotron sometimes with and sometimes without the addition of a
blackbody. This motivates our choice of the synchrotron+blackbody
photon model for fitting the spectra of the GRB~\textit{141028A}. We
note that not all GRB spectra are compatible with this photon
model. There exist GRBs such as GRB~\textit{090902B} with a clear
dominant photospheric component originating from subphotospheric
dissipation \citep{Peer:2005,Ryde:2010}.

The synchrotron model implemented is fully physical in the sense that
the spectral shape comes from the synchrotron emissivity and electron
distribution alone. The spectral fitting \textit{should not} be
confused with fitting the empirical Band function \citep{Band:1993}
with its variable low-energy index ($\alpha$). The blackbody is only
included in time bins that statistically require it via a likelihood
ratio test. We find that this statistical requirement produces a
continuous presence of the blackbody for the first part of the
observation. The data are well fit by the two-component model (see
Table \ref{tab:gbmanal} for a summary of the spectral fits) and we
observe two key features that motivate our investigation of the GRB
under the framework of the external shock model: $\Ep$ evolves as a
broken power law in time and the blackbody is only significant for a
duration of $\sim$10 seconds compared to the total prompt duration of
$\sim$25~s (see Section~\ref{sec:photosphere}).

\begin{deluxetable*}{rrccccc}
\tablecaption{Results of the time-resolved spectral analysis of GBM+LLE data (10 keV - 300 MeV).}
\tablecolumns{7}
\tablewidth{0pt}
\tablehead{\colhead{T$_{\rm start}$ [s]} & \colhead{T$_{\rm stop}$ [s]} & \colhead{$F_{\rm E}^{\synch}$ [erg s$^{-1}$]\tablenotemark{b}}  & \colhead{$\Ep$ [keV]\tablenotemark{c}} & \colhead{$p$\tablenotemark{d}} & \colhead{$F_{\rm E}^{\BB}$ [erg s$^{-1}$]} & \colhead{$kT$ [keV]}}
\startdata
6.0                                      & 7.0                          & $(1.79\pm0.34)\cdot 10^{-6}$                  & $1109\pm336$          & $10.00$\tablenotemark{a}               & $(1.61\pm3.43)\cdot 10^{-8}$               & $24.55\pm26.48$ \\
7.0                                      & 8.0                          & $(3.19\pm0.54)\cdot 10^{-6}$                  & $1378\pm472$          & $5.74\pm1.21$                          & $(1.04\pm0.54)\cdot 10^{-7}$               & $25.12\pm4.50$  \\
8.0                                      & 9.0                          & $(4.99\pm0.59)\cdot 10^{-6}$                  & $1383\pm310$          & $5.21\pm0.58$                          & $(2.76\pm0.59)\cdot 10^{-7}$               & $40.69\pm5.40$  \\
9.0                                      & 10.0                         & $(7.25\pm0.68)\cdot 10^{-6}$                  & $1504\pm310$          & $4.59\pm0.30$                          & $(1.91\pm0.68)\cdot 10^{-7}$               & $29.98\pm3.94$  \\
10.0                                     & 11.0                         & $(6.38\pm0.58)\cdot 10^{-6}$                  & $738\pm108$           & $4.11\pm0.18$                          & $(4.38\pm0.58)\cdot 10^{-7}$               & $60.01\pm9.61$  \\
11.0                                     & 12.0                         & $(5.40\pm0.42)\cdot 10^{-6}$                  & $572\pm79$            & $4.18\pm0.21$                          & $(3.21\pm0.42)\cdot 10^{-7}$               & $35.78\pm4.54$  \\
12.0                                     & 13.0                         & $(6.26\pm0.60)\cdot 10^{-6}$                  & $498\pm71$            & $4.09\pm0.20$                          & $(6.96\pm0.60)\cdot 10^{-7}$               & $73.11\pm10.11$ \\
13.0                                     & 14.0                         & $(6.56\pm0.42)\cdot 10^{-6}$                  & $520\pm55$            & $4.23\pm0.18$                          & $(3.12\pm0.42)\cdot 10^{-7}$               & $37.28\pm5.62$  \\
14.0                                     & 15.0                         & $(4.48\pm0.31)\cdot 10^{-6}$                  & $310\pm46$            & $3.67\pm0.13$                          & $(1.33\pm0.31)\cdot 10^{-7}$               & $21.56\pm3.78$  \\
15.0                                     & 16.0                         & $(5.14\pm0.35)\cdot 10^{-6}$                  & $288\pm34$            & $3.65\pm0.12$                          & $(2.13\pm0.35)\cdot 10^{-7}$               & $31.06\pm5.68$  \\
16.0                                     & 17.0                         & $(2.90\pm0.23)\cdot 10^{-6}$                  & $168\pm16$            & $3.69\pm0.16$                          & \nodata                                    & \nodata         \\
17.0                                     & 18.0                         & $(2.48\pm0.21)\cdot 10^{-6}$                  & $175\pm21$            & $3.52\pm0.14$                          & \nodata                                    & \nodata         \\
18.0                                     & 19.0                         & $(1.75\pm0.20)\cdot 10^{-6}$                  & $147\pm20$            & $3.68\pm0.22$                          & \nodata                                    & \nodata         \\
19.0                                     & 20.0                         & $(2.03\pm0.19)\cdot 10^{-6}$                  & $123\pm18$            & $3.24\pm0.11$                          & \nodata                                    & \nodata         \\
20.0                                     & 21.0                         & $(1.77\pm0.18)\cdot 10^{-6}$                  & $122\pm18$            & $3.33\pm0.13$                          & \nodata                                    & \nodata         \\
21.0                                     & 22.0                         & $(1.42\pm0.19)\cdot 10^{-6}$                  & $148\pm23$            & $3.77\pm0.28$                          & \nodata                                    & \nodata         \\
22.0                                     & 23.0                         & $(1.30\pm0.17)\cdot 10^{-6}$                  & $127\pm23$            & $3.43\pm0.18$                          & \nodata                                    & \nodata         \\
23.0                                     & 24.0                         & $(1.10\pm0.16)\cdot 10^{-6}$                  & $87\pm20$             & $3.20\pm0.16$                          & \nodata                                    & \nodata         \\
24.0                                     & 25.0                         & $(6.91\pm1.65)\cdot 10^{-7}$                  & $105\pm28$            & $3.64\pm0.44$                          & \nodata                                    & \nodata         \\
25.0                                     & 26.0                         & $(9.28\pm1.44)\cdot 10^{-7}$                  & $139\pm37$            & $3.35\pm0.19$                          & \nodata                                    & \nodata         \\
26.0                                     & 27.0                         & $(6.06\pm1.34)\cdot 10^{-7}$                  & $152\pm44$            & $3.86\pm0.48$                          & \nodata                                    & \nodata         \\
27.0                                     & 28.0                         & $(7.34\pm1.39)\cdot 10^{-7}$                  & $126\pm44$            & $3.27\pm0.22$                          & \nodata                                    & \nodata         \\
\enddata
\tablenotetext{a}{fixed}
\tablenotetext{b}{synchrotron energy flux}
\tablenotetext{c}{synchrotron $\vFv$ peak}
\tablenotetext{d}{$e^-$ spectral index}
\label{tab:gbmanal}
\end{deluxetable*}

We calculate the total k-corrected energy \citep{Hogg:2002} in the
synchrotron component in the 30 keV-300 MeV interval by summation over
each time bin: $E_{\synch} = 4 \pi d_{\rm L}^2 \sum F^{\synch} \Delta
t_{\rm i} \sim \Eiso\; \text{erg}$.  The total isotropic energy of the
burst is estimated to be several times larger. In the following, we
take the total isotropic energy (kinetic + radiative) of the
blast-wave to be $E_{\rm iso}\cong 10^{55}\text{erg}$ while noting
that this is an extremely high value. Nevertheless, the GRB is
extremely bright and no mechanism is known that is efficient enough to
convert the entire rest mass of the progenitor to radiation.

For the late time GeV emission, we performed an unbinned likelihood
analysis of the LAT data to recover the energy flux ($F_{\rm E}$) and
photon index ($\gamma_{\rm ph}$) of the emission (see Table
\ref{tab:LATSpectral}) with the \texttt{gtlike} program distributed
with the \Fermi \texttt{ScienceTools}\footnote{We used version
  09-34-02 available from the \Fermi Science Support Center
  \url{http://fermi.gsfc.nasa.gov/ssc/}}.  We selected
\texttt{P7REP\_SOURCE\_V15} photon events from a 15$^{\circ}$ circular
region centered at the Swift XRT position (R.A.=322$\fdg$60,
Dec.=$-$0$\fdg$23, J2000) and within 105\de from the local zenith (to
reduce contamination from the Earth limb). Events with measured energy
from 100 MeV to 10 GeV are included in our analysis (the highest
energy event associated with this GRB has an energy of 3.8 GeV and
arrives 157.5 seconds after the GBM trigger time). Further details on
the LAT analysis are discussed in Appendix \ref{sec:latinfo}.

\begin{deluxetable*}{rrccc}
\tablecaption{Results of the time-resolved spectral analysis of LAT data  (100 MeV - 10 GeV) where TS is the value of the Test Statistic.}
\tablecolumns{5}
\tablewidth{0pt}
\tablehead{
 \colhead{T$_{\rm start}$ [s]} & \colhead{T$_{\rm stop}$ [s]} & \colhead{TS}  &  \colhead{$\gamma_{\rm ph}$}  &  \colhead{$F_{\rm E}$ [erg s$^{-1}$cm$^{-2}$]}
}
\startdata
13.3   & 23.7    & 26.3  & $-$2.9$\pm$0.7 & (7.4$\pm$3.5)$\times$10$^{-8}$ \\
23.7   & 75.0    & 26.8  & $-$2.1$\pm$0.5 & (1.6$\pm$1.0)$\times$10$^{-8}$ \\
75.0   & 237.1   & 32.9  & $-$1.5$\pm$0.4 & (7.0$\pm$4.5)$\times$10$^{-9}$ \\
316.2  & 724.8   & 7.9   & \nodata        & $<$2.0$\times$10$^{-8}$        \\
3585.8 & 6446.8  & 1.8   & \nodata        & $<$7.7$\times$10$^{-10}$       \\
9307.8 & 10000.0 & 0.0   & \nodata        & $<$3.1$\times$10$^{-9}$        \\
\enddata

\label{tab:LATSpectral}

\end{deluxetable*}

\section{External Shock Analysis}
\label{sec:esa}
\subsection{$\Ep$ Evolution}
\label{sec:epevo}
\begin{figure}[h]
  \centering

 \includegraphics[scale=1]{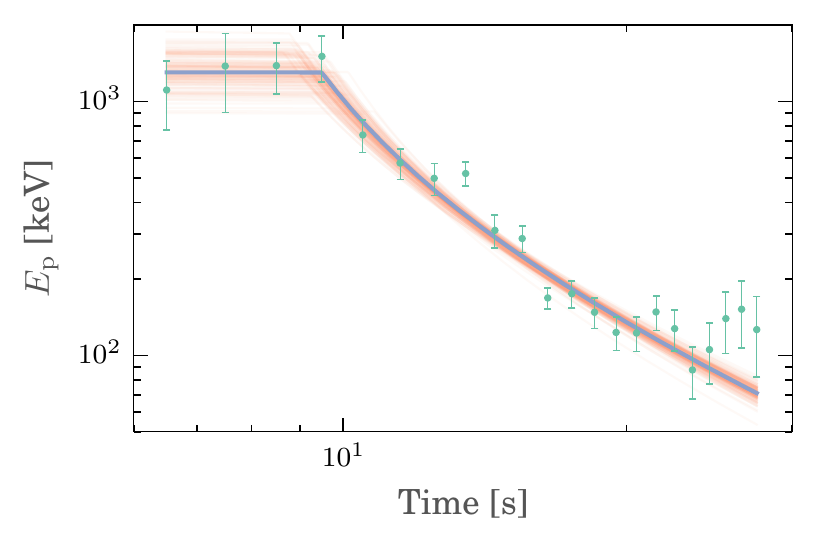}

 \caption{Bayesian fit of the $\Ep$ evolution of
   GRB~\textit{141028A}. The maximum-likelihood point is indicated by
   the {\it blue} curve and posterior samples are in {\it orange}. The
   fitted parameters are indicated in Table~\ref{tab:epfit1}.}
  \label{fig:fit}
\end{figure}

After performing spectral fits to the data we can test the external
shock model by fitting the evolution of the recovered synchrotron
$\Ep$ with Equation~\ref{eq:ep}. The fit is performed with a Bayesian
analysis tool built upon the {\tt MULTINEST} \citep{Feroz:2009}
software which allows us to fully explore the correlated parameter
space of the model. The free parameters in the fit are $\G0$, $\eta$,
$g$, and $\q3$ to which we assign flat priors that are consistent with
physical expectations ($\G0\in\{ 10,1500 \}$, $\eta\in\{ 0,2 \}$
$g\in\{ 0,3 \}$ , $\q3 \in\{ 0,10 \}$). We also note that it is not
possible to fit for the density $n_0$, which we set at $n_0 =
\{1,10,100\} \; \text{cm}^{-3}$ with different values mainly affecting
the value of $\G0$ recovered. Calculations in the text and figures
assume $n_0=100$ cm$^{-3}$ but are complemented with calculations at
the other values in associated tables. Figure \ref{fig:fit} shows the
fit of the $\Ep$ evolution and the best-fit parameters are detailed in
Table~\ref{tab:epfit1}.

\begin{deluxetable*}{r c c c c c}
\tablecaption{The best-fit parameters of the $\Ep$ evolution fit for each assumed value of $n_0$ as well as the inferred $r_{\rm d}$.\label{tab:epfit1}}
\tablecolumns{6}
\tablewidth{0pt}
\tablehead{
 \colhead{$n_{0}$ [cm$^{-3}$]} & \colhead{$\Gamma_0$} & \colhead{$\eta$} & \colhead{$q_{-3}$} & \colhead{$g$} & \colhead{$r_{\rm d}$ [cm]}
}
\startdata
 1   & $1125.9^{+16.6}_{-14.3}$ & $0.00^{+0.08}_{-0.00}$ & $0.17^{+0.01}_{-0.13}$ & $1.25^{+0.1}_{-0.1}$ & $1.1\cdot 10^{17} $  \\[1mm]
 10  & $844.0^{+12.3}_{-10.9}$   & $0.00^{+0.08}_{-0.00}$ & $0.17^{+0.02}_{-0.14}$ & $1.26^{+0.1}_{-0.11}$  & $ 6.0\cdot 10^{16} $ \\[1mm]
 100 & $632.16^{+7.7}_{-4.5}$   & $0.03^{+0.05}_{-0.02}$ & $0.17^{+0.05}_{-0.06}$  & $1.26^{+0.09}_{-0.11}$ & $ 3.4\cdot 10^{16} $ \\
\enddata
\end{deluxetable*}

\begin{figure}[h]
  \centering
  \includegraphics{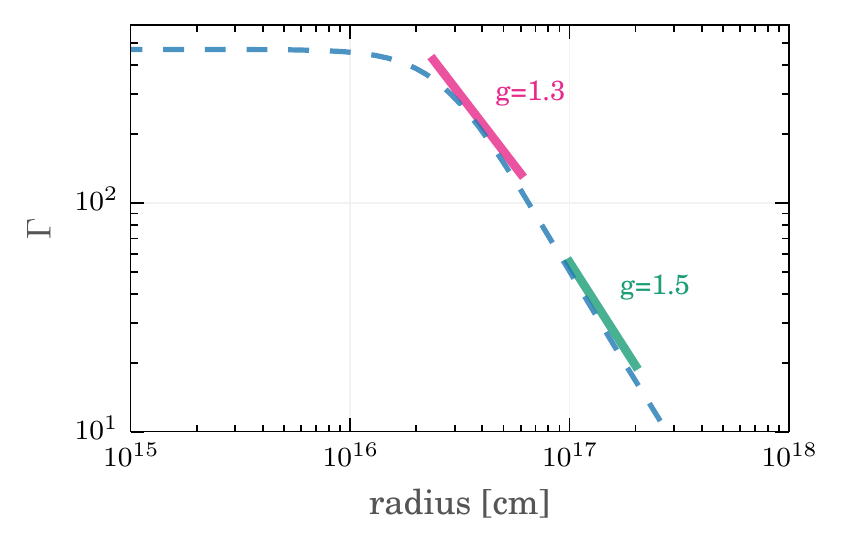}
  \caption{The blast-wave Lorentz factor evolution from our numerical
    simulation of the external shock model (\emph{blue dashed
      line}). We illustrate that we are most likely finding the
    blast-wave still evolving in the transition phase, which accounts
    for the shallow value of g found in the fit of $\Ep$.}
  \label{fig:num}
\end{figure}

We find that $g\simeq1.3$, which is less than what is expected if the
blast-wave decelerates adiabatically ($g=1.5$). This suggests that the
blast-wave is still transitioning to the asymptotic limits of Equation
\ref{eq:gamma}. To test this assumption, we numerically solved for the
evolution of $\Gamma$ with radius via the equations of energy and
momentum conservation and examined the transition phase in Figure
\ref{fig:num}. There is clearly a region that corresponds to our
recovered value of $g\simeq1.3$. Late (hundreds of seconds after the
trigger) time observations smoothly connecting the prompt and
afterglow emission would allow us to measure the asymptotic value of
$g$. Yet, with the recovered fit parameters, we can calculate several
physical properties of the outflow including $r_{\rm d} = \xd$~cm and
the resulting predicted $P_p(t)$ evolution though we will assume
$g\simeq 1.6$ to calculate fluxes as this is the value reached
asymptotically in the near adiabatic case \cite{Chiang:1999}. These
parameter values are used in the following sections to gain more
insight about the GRB and its surrounding environment.

\subsection{Peak Flux - $\Ep$ plane}
\label{sec:eppp}
\begin{figure}[h]
  \centering
  \includegraphics{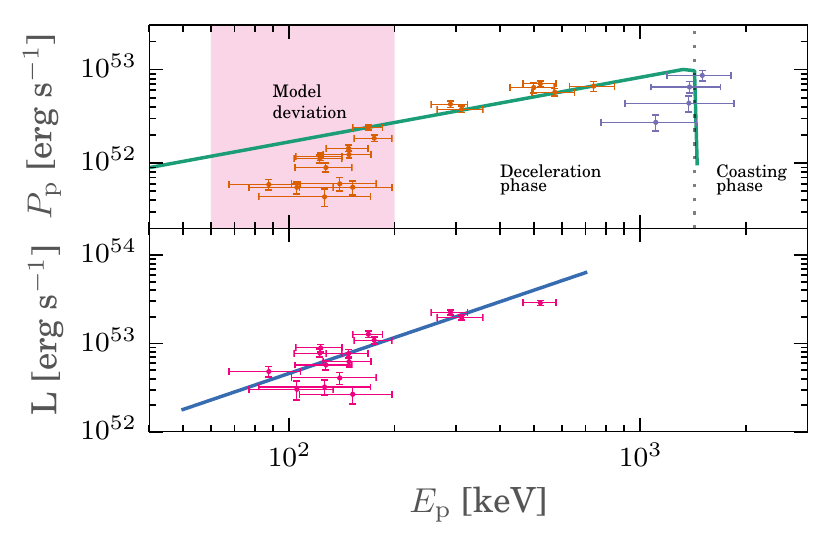}
  \caption{The $P_{\rm p} - \Ep$ plane of GRB~\textit{141028A}
    (\textit{top}) where $P_{\rm p}$ is the $\vFv$ flux calculated at
    $\Ep$. The predicted evolution ({\it green}) is produced by
    substituting the recovered parameters of the $\Ep$ evolution fit
    into Equation ~\ref{eq:pt}. The flux given by Equation \ref{eq:pt}
    is scaled by a factor of 9 to match the data.  At late times (pink
    region), the observed flux decays faster than the model
    predicts. The decay phase synchrotron HIC ({\it blue})
    (\textit{bottom}). The fitted value of the slope is $\zeta=1.33\pm
    0.22$. The decay phase is selected as the portion of the
    lightcurve that monotonically decreases with time i.e. from 10-27
    s.}
  \label{fig:cor}
\end{figure}

Using the recovered parameters from the fit, we can use Equations
\ref{eq:ep} and~\ref{eq:pt} to predict the evolution in the $P_{\rm
  p}$-$\Ep$ plane. In Figure~\ref{fig:cor}, the data from
GRB~\textit{141028A} is plotted with the predicted curve from the
external shock model derived from the $\Ep$ fits. Noting the
discussion in Section \ref{sec:epevo}, negative fluxes would be
obtained for $g<1.5$ in Equation \ref{eq:p0}. Since the obtained value
of $g$ via the fit to the $\Ep$ evolution is not actually measuring
the asymptotic behavior of the blast-wave evolution as is intended, we
set $g=1.6$ in Equation \ref{eq:p0} corresponding to a nearly
adiabatic blast-wave motivated by the low magnetic fields measured via
the spectra and late time XRT observations (see Section
\ref{sec:afterglow} for full details). This choice of $g$ does not affect the other
parameters in the fit as $g$ only modifies the late time evolution of
$\Ep$ (see Figure \ref{fig:demo}).

The analytic model given by Equation \ref{eq:pt} and Equation
\ref{eq:p0} is consistent with the data up to a scale factor of 9
which can easily be due to insufficient knowledge of all intrinsic
parameters, on the simplified evolution law for the flux, and on the
assumed value of $g$. It is pertinent to note that the parameters used
in the $P_{\rm p}(t)$ prediction come from the fitted $\Ep$ evolution
alone. In addition, the predicted $P_{\rm p}(t)$ is independent of the
observed flux values in the data. The close agreement between the
predicted and measured $P_{\rm p}(t)$ values is evidence that the
observed flux at the peak and the observed $\vFv$ peak energy are
linked in a manner predicted by the external shock model. We do note
the deviation from the predicted curve at late times. This therefore
provides strong evidence that these two independent quantities are
linked in a manner predicted by the external shock model.

\subsection{The Luminosity - $\Ep$ plane}

The commonly observed correlation in the decay phase of many GRB
pulses between luminosity and $\Ep$ in the form,
\begin{equation}
  \label{eq:hic}
  L\propto\Ep^{\;\zeta}{\rm ,} 
\end{equation}

\noindent sometimes referred to as the hardness-intensity correlation
(HIC) \citep{Golenetskii:1983}, is potentially another clue to the
radiation mechanism responsible for the observed emission. It is quite
simple to relate the luminosity and $\vFv$ peak energy of many common
radiation mechanisms analytically. However, the spread in HIC power
law slopes observed across many GRBs is difficult to explain with one
mechanism \citep{Borgonovo:2001}. However, single pulse GRBs fit with
a synchrotron photon model have been found to have HIC power law
slopes of $\sim1.5$ without much spread \citep{Burgess:2014}.

In \citet{Dermer:2004}, an analytic form for the HIC resulting from
synchrotron emission is derived and parameterized to account for
various effects including expansion geometry, magnetic flux-freezing,
and radiative regime. Following this derivation, we can make the
following predictions for the evolution of the important quantities in
determining the relationship between $L$ and $\Ep$ during the
deceleration phase of a blast-wave.  We have $ x\propto t^{1/(2g+1)}$,
$\Gamma \propto t^{-g/(2g+1)}$, $ \Ep \propto \Gamma B \gamma_{\rm
  min}^2$, and $L \propto \Gamma^2 B^2 \gamma_{\rm min}^2$. Here
$\gamma_{\rm min}$ is the minimum electron energy in an assumed shock
accelerated power law distribution (though the result is insensitive
to the fact that we assume a power law) and $B$ is the strength of the
magnetic field. There are two regimes of cooling for electrons: fast
and slow \citep{Sari:1998}. Each regime can be characterized by how
$\gamma_{\rm min}$ evolves with time such that
\begin{equation}
\label{eq:gel}
\gamma_{\rm min}\propto\left\{
\begin{array}{ll}
 \Gamma^4\propto t^{-4g/(2g+1)}	& {\rm slow \; cooling} \\
  (x\Gamma )^{-1}\propto t^{-2g/(2g+1)}	& {\rm fast \; cooling}
\end{array}
\right. .
\end{equation}
Since $L\propto B \Gamma \Ep$, we can write
\begin{equation}
\label{eq:gel}
L\propto\left\{
\begin{array}{ll}
 \Ep^{\nicefrac{3}{2}}	& {\rm slow \; cooling} \\
 \Ep^{1+g}	& {\rm fast \; cooling}
\end{array}
\right.
\end{equation}

\noindent When fitting the $L-\Ep$ data in the decay phase (10-27 s)
we find $\zeta=1.33\pm 0.22$ (see Figure~\ref{fig:cor}). This is
closer to the value expected for the slow-cooling regime which falls
in line with the use of slow-cooling synchrotron to fit the
time-resolved spectra. An a posteriori justification for the use of
slow-cooling synchrotron to fit the spectra is discussed in Section
\ref{sec:afterglow}.

\subsection{The photosphere}
\label{sec:photosphere}

We interpret the observed thermal component in the framework of
photospheric emission, i.e., when the outflow becomes transparent at a
radius $\sim 10^{10} - 10^{12}$ cm, i.e., much below $r_{\rm
  d}$. Because of highly relativistic motion, the observed time delay
between photons from the photosphere and those from the external shock
is small. As the thermal component is not dominant, the errors in the
calculation of outflow parameters derived in \citet{Peer:2007} are
large and we therefore did not use this formalism herein.

We note that the blackbody is not statistically significant (below
3-$\sigma$ confidence level) in the first five seconds after
trigger. However, photons from the photosphere are expected to arrive
slightly before synchrotron photons. The delay between the
photospheric photons and those at the peak of the external shock light
curve can be estimated as $(r_{\rm d} - r_{\rm ph})/ c \Gamma_0 \simeq
9.5$ s corresponding roughly $t_{\rm d}$. Thus, we added five seconds
to the duration of the photospheric emission. The identification of
the photospheric emission can give important constraints.  First, the
duration of the photospheric component ($\Delta t_{\text{ph}}$) sets
the width of the expanding outflow $(W)$ at the photosphere
\citep[see][]{Begue:2013,Begue:2014,Vereshchagin:2014},

\begin{equation}
\frac{\Delta t_{\text{ph}}}{1+z} = \frac{R_{ \text{ph}}}{2 \Gamma_0^2 c} + \frac{W}{c} + \frac{R_{ \text{ph}}}{2 \Gamma_0^2 c}
\end{equation}

\noindent where the first term accounts for the expansion time of the outflow
up to the photosphere, the second term is the light-crossing time of
the outflow and the last term is the angular timescale at the
photosphere. For the observed duration at hand and for the Lorentz
factor values expected in GRB physics, the second term in the
right-hand side of the equation dominates. It implies $W \sim c \Delta
t_{\text{ph}} /(1+z) \sim 1.35 \cdot 10^{11} \text{cm}$.

As the width $W$, the Lorentz factor $\Gamma$ and the total energy of
the blast wave are constrained, the state of the reverse shock can be
a posteriori studied. Following the discussion in \citet{dermerbook},
we define

\begin{equation}
\xi  \sim \frac{54}{\left ( \frac{\Gamma}{300} \right )^{\nicefrac{4}{3}} W_8^{\nicefrac{1}{2}} } \left (  \frac{E_{52}}{n_0} \right )^{\nicefrac{1}{6}}  \text{,}
\end{equation}

\noindent such that $x_{\rm NR}=\xi r_{\rm d}$ is the radius at which
the reverse shock becomes relativistic. We find $\xi=0.8$ which
implies that the reverse shock crosses the expanding outflow while it
is only mildly relativistic. This justifies the use of
Equation~\ref{eq:gamma} in our treatment. We note the value of $\xi$
is the same for each combination of $\G0$ and $n_0$ (see Table
\ref{tab:eb}).

\section{Discussion}

\subsection{Compatibility with the Late-time X-ray/optical Emission}
\label{sec:afterglow}
\begin{figure}[h]
  \centering
  \includegraphics{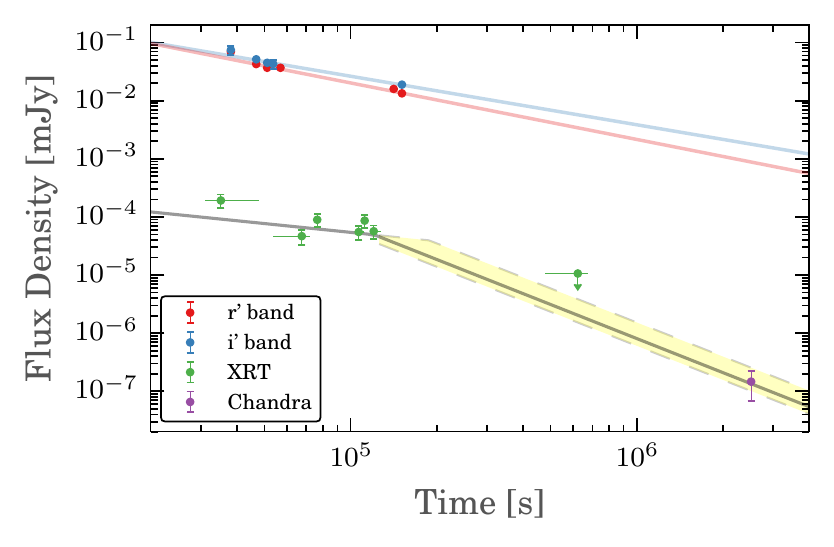}
  \caption{The X-ray and optical afterglow of GRB~\textit{141028A}. A
    power law fit best describes the optical data while a broken power
    law fit, indicative of a jet break, best describes the X-ray
    data. The \emph{yellow} shaded region indicates the 90\%
    uncertainty in the break time due to the sparse sampling of data
    prior to the Chandra follow-up observation.}
  \label{fig:chandraxrt}
\end{figure}

Under the assumption of an external shock in the prompt phase, the
afterglow is produced by the same process (that is the deceleration of
a blast wave by the CBM), and we do not expect any break in the light
curve, as already seen in several GRBs observed by $Swift$
\citep{Zhang:2006}.  Unfortunately, the XRT observations only cover a
short time range beginning a $\text{few} \sim10^{4}$ s after the
prompt emission (See Figure \ref{fig:chandraxrt}). At most, we can say
that XRT late time observations are compatible with the emission from
a decelerating blast-wave in the semi-radiative regime, i.e., a region
between $1.5 < g < 3$.

In addition, combining observations by XRT and Chandra, displayed in
Figure~\ref{fig:chandraxrt}, a break can be identified in the light
curve. The exact time of the break is unconstrained but it ranges from
$10^5\; \text{s} < t_{\rm b} < 4.3 \cdot 10^{5}\; \text{s}$.
Identifying the break as a jet break, the jet opening angle can be
estimated to be \citep{Ghirlanda:2004}

\begin{equation}
  \theta_{\rm j}  = 0.161 \left ( \frac{t_{\rm b}}{1+z} \right )^{\nicefrac{3}{8}} \left ( \frac{\eta_{\gamma}n_0 }{E_{\rm \gamma,iso,52}} \right )^{1/8} 
\end{equation}

\noindent where $t_{\rm b}$ is the break time in days and
$\eta_{\gamma}$ is the radiative efficiency taken to be 0.1
corresponding to our assumption of $E_0 = 10^{55}$ erg. Thus, the
opening angle ranges from $2.65^{\circ} < \theta_{\rm j} <
8.15^{\circ}$, when considering the fitted parameters obtained in
Section \ref{sec:epevo}. Therefore, the radiative energy budget of the
burst is considerably reduced to few $\sim 10^{51}$~erg (see Table
\ref{tab:beam}).

\begin{deluxetable}{r c c c  c}
\tablecaption{Values of $\epsilon_{\rm B}$ from the afterglow and $\Ep$ assuming different values of $n_{0}$ as well as the location of the relativistic reverse shock ($\xi$).\label{tab:eb}}
\tablecolumns{5}
\tablewidth{0pt}
\tablehead{\colhead{$n_0$} & \colhead{$\G0$} & \colhead{$\epsilon_{\rm B}$ via $\nu_{\rm c}$} & \colhead{$\epsilon_{\rm B}$ via $\Ep$} & \colhead{$\xi$}}
\startdata
1                       & 1125.9           & $<1.2\cdot 10^{-3}$                            & $\sim 2.9\cdot 10^{-9}$                 & 0.8 \\
10                       & 884.0           & $<2.7\cdot 10^{-4}$                            & $\sim 2.9\cdot 10^{-9}$                 & 0.8 \\
100                      & 632.2           & $<5.7\cdot 10^{-5}$                            & $\sim 2.9\cdot 10^{-9}$                 & 0.8 \\
\enddata
\end{deluxetable}

\begin{deluxetable}{lccc}
\tablecaption{Corrected Emitted Burst Energy ($E_{\rm cor}$)}
\tablecolumns{4}
\tablewidth{0pt}
\tablehead{\colhead{$n_0$ [cm$^{-3}$]} & \colhead{$t_{\rm b}$ (s)} & \colhead{$\theta_{\rm j}$ (degrees)} & \colhead{$E_{\rm cor}$ (erg)} }

\startdata
\multirow{2}{*}{1} & $1 \cdot10^5$ & 2.65 & $1.0\cdot 10^{51}$\\
&$4.3 \cdot 10^5$ & 4.58  & $2.9\cdot 10^{51}$\\[3mm]
\multirow{2}{*}{10} &$1 \cdot10^5$ & 3.53 & $1.7\cdot 10^{51}$\\
&$4.3 \cdot 10^5$ & 6.1  & $5.2\cdot 10^{51}$\\[3mm]
\multirow{2}{*}{100} &$1 \cdot10^5$ & 4.71 & $3.1\cdot 10^{51}$\\
&$4.3 \cdot 10^5$ & 8.15  & $9.2\cdot 10^{51}$\\
\enddata

\label{tab:beam}
\end{deluxetable}

Finally, additional constraints can be obtained from simultaneous
optical and X-ray observations at around 50 ks after the trigger. The
spectrum is consistent with a single power-law of index $\beta = 1.29
\pm 0.07$. Therefore, we can deduce that the cooling frequency is
above the XRT frequency: $\nu_{\rm c} > \nu_{\XRT}$. From the
expansion of a blast-wave in a constant density CBM, $\nu_{\rm c}$ can
be estimated as \citep{Panaitescu:2000}

\begin{equation}
  \nu_{\rm c} = 3.7 \cdot 10^{16} E_{53}^{-1/2} n_0^{-1} (Y+1)^{-2} \epsilon_{\rm B,-2}^{-3/2} T_{\rm d}^{-1/2}\; {\rm Hz,}
\end{equation}

\noindent where $\epsilon_{\rm B}$ parameterizes the magnetic field in
the shocked ISM, $T_{\rm d}$ is the time in days and $Y$ is the
Compton parameter, that we chose to be zero for simplicity. This leads
to an upper limit $\epsilon_{\rm B} < 2.3 \cdot 10^{-5}$ taking
$T_{\rm d}=1$ day and assuming $n_0 = 100$ and $E_{53}=100$ which
implies $\G0=632$ (see however, Table \ref{tab:eb} for values
corresponding to different parameter choices).

Also, this value can be cross-checked by considering the peak energy in
the first seconds of the prompt phase, which can be evaluated as:

\begin{equation}
  \Ep \simeq \frac{\sqrt{32 \pi \epsilon_{\rm B} m_{\rm p} c^2 n_0 \Gamma^2}}{B_{\rm crit}}\Gamma \gamma_{\rm min}^2 
\end{equation}

\noindent where $B_{\rm crit} = 4.414\cdot 10^{13} \text{G}$ is the
critical magnetic field, $\kappa$ parameterizes the minimum Lorentz
factor $\gamma_{\rm min}$ of the accelerated electrons such that
$\gamma_{\rm min} = \kappa \Gamma (\nicefrac{m_{\rm p}}{m_{\rm e}})$,
$m_{\rm e}$ is the mass of an electron, and $\Ep$ is given in units of
the mass of an electron. Using $\Ep \sim 1.3$ MeV and $n_0=100$ as
obtained from the data, and assuming $\kappa=1$ gives
$\epsilon_{\rm B} \sim 3.0\cdot 10^{-9} $ (see also Table
\ref{tab:eb}). This value is consistent with the upper limit obtained
from the late afterglow\footnote{ However we note that this value is
  very sensitive to $\kappa$ which is fairly unconstrained: as an
  example, with $\kappa = 0.1$, it becomes $\epsilon_{\rm B} \sim
  2.8\cdot 10^{-5} $, incompatible with the upper limit obtained from
  the afterglow.}.

Finally, the cooling time of an electron with Lorentz factor
$\gamma_{\rm min}$ can be compared to the expansion time of the
blast-wave at the observed luminosity peak which is on the order of
the $t_{\rm d}$. We find that they are comparable, leading to
efficient energy extraction from the electrons, without drastically
changing the electron distribution function, \textit{i.e.}, creating
an additional power law over several orders of magnitude at low-energy
characteristic of fast-cooled (or completely cooled) electrons.

\subsection{LAT late time GeV emission}
\label{sec:lat}

$Fermi$ LAT observed a GeV component over the duration of $10-10^3$
s. Unfortunately, there were no simultaneous observations at other
wavelengths from which a broadband spectrum could be
obtained. Nevertheless, we can use the analytic flux evolution derived
in \citet{Dermer:1999} to check if the observed LAT fluxes are
compatible with the model. We must first address a few caveats: the
analytic model assumes no synchrotron self-Compton (SSC) emission and
the fluxes are sensitive to our lack of knowledge about the total
burst energy (which we assume to be $10^{55}$ erg for the purpose of
calculation) as well as the degeneracies in the recovered parameters
from the $\Ep$ evolution fit.

In the past years, a debate concerning the origin of the late time
(few seconds after the trigger) high-energy LAT emission has taken
place, opposing a synchrotron mechanism \citep{Kumar:2009,Kumar:2010},
SSC \citep{Panaitescu:2000a}, and pair-loading of the CBM
\citep{Beloborodov:2005} as the possible candidates for this
emission. First, we can assess the SSC component, and show that with
the parameters at hand from the prompt emission the SSC flux in the
LAT bandpass is much less than the synchrotron flux. We can constrain
the SSC flux in the LAT bandpass by following the discussion in
\citet{Beniamini:2013} (see Appendix \ref{sec:sscCalc} for further
details). Via Equation \ref{eq:fracit2}, the ratio of SSC flux to
synchrotron flux in the LAT bandpass ($\mathcal{F}_{\SSC}$) can be
computed by substituting values derived in Section \ref{sec:afterglow}
to yield $\mathcal{F}_{\SSC} \simeq 10^{-8}$. The numerical result shows
that the SSC flux in the LAT band is much less than the synchrotron
flux. However, the computation \textit{relies} on the assumption that
electrons are fast cooled either by synchrotron or by inverse Compton
of the produced synchrotron photons, that is to say that all the energy
is emitted by one or the other mechanism. It comes with two
consequences:
\begin{enumerate}
\item the Compton parameter \textit{is} an upper limit as shown in
  \citet{Sari:1996}. Therefore the result of Eq.(\ref{eq:fracit2}) is
  only an upper limit,
\item as a result of the assumption of fast cooling, the electron
  distribution function extends to small Lorentz factors. These
  electrons are not in the Klein-Nishina regime and can efficiently
  upscattered synchrotron photon in the LAT band. However, we found
  that electrons with Lorentz factor $\gamma_{\rm min}$ are not
  strongly cooled over a dynamical time. Therefore the electron
  distribution function does not extend substantially below
  $\gamma_{\rm min}$, reducing the inverse Compton flux.
\end{enumerate}

\begin{figure}[h]
  \centering
  \includegraphics{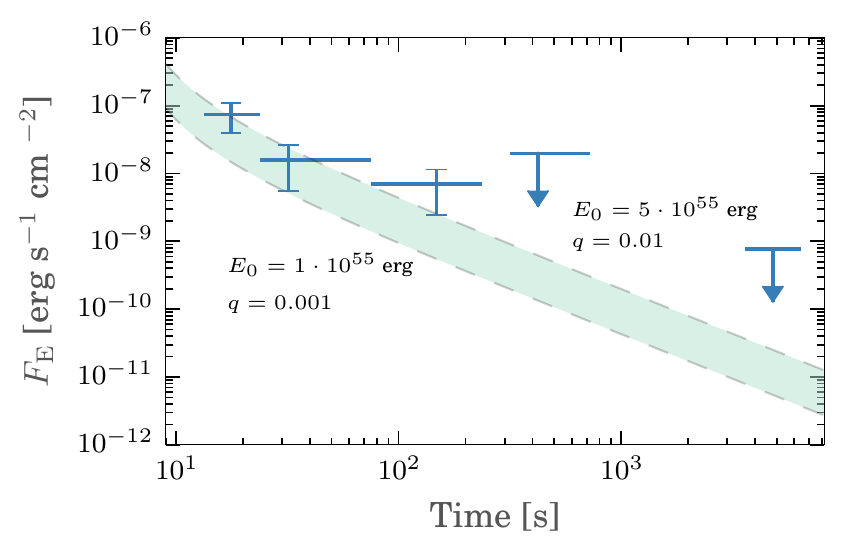}
  \caption{The LAT GeV integrated energy fluxes (100 MeV - 10 GeV) are
    compared with a range of expected fluxes from the analytic model
    of \citet{Dermer:1999} that includes only synchrotron
    emission. Suitable agreement can be found for a valid range of
    assumed blast-wave parameters.}
  \label{fig:latflux}
\end{figure}

\begin{figure}[h]
  \centering
  \includegraphics{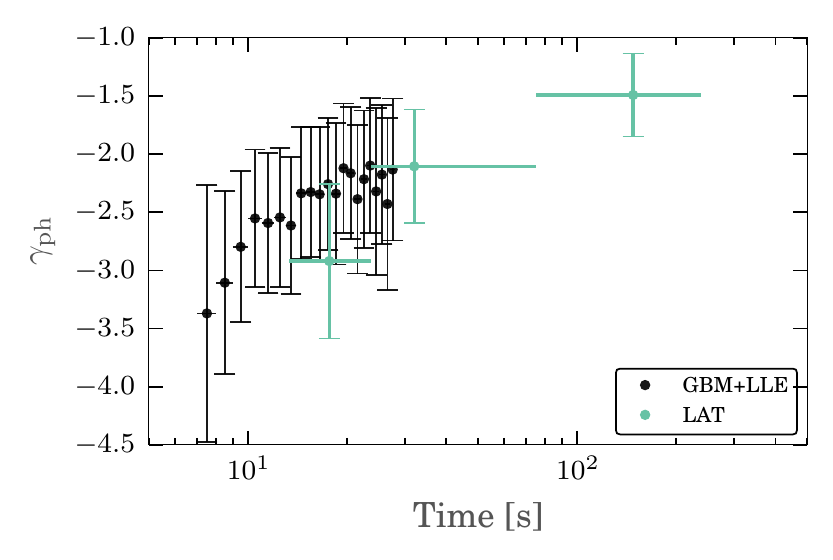}
  \caption{The photon indices from the GBM+LLE fits and LAT fits. A
    clear evolution is seen into the LAT band pass with the indices
    all in line with what could be expected from synchrotron-emitting
    high-energy electrons accelerated into a power law momentum-distribution.}
  \label{fig:indx}
\end{figure}

We now consider that the GeV emission is the result of synchrotron
emission from the same forward-shock responsible for the lower energy
prompt phase. Using the analytic estimate for the blast-wave
evolution, we estimate the expected flux of the synchrotron emitting
blast-wave and compare it to the integrated (100 MeV - 10 GeV) LAT
energy flux ($F_{\rm E}$) by integrating Equation 1 from
\citet{Dermer:1999} across the LAT bandpass (see Figure
\ref{fig:latflux}). Because of our lack of knowledge about the
intrinsic total energy of the blast-wave and the degeneracies in the
fit parameters ($q$, $n_0$), we vary the parameters across a broad
range and find that the model gives consistent limits within an order
of magnitude. We also note that this model is simple and will not be
as accurate as a full numerical solution. Additionally, using the
power-law electron index ($p$) found from GBM+LLE synchrotron fits, we
compute the synchrotron photon indices via the transformation
$\gamma_{\rm ph} = -(p+1)/2$ and compare them with the photon indices
found in the LAT alone (Figure \ref{fig:indx}). There is a clear
evolution of the synchrotron photon index increasing that smoothly
transitions to the indices observed by the LAT.

Therefore, we conclude that the high energy emission from the LAT is
compatible with synchrotron emission alone, and does not require an
additional mechanism. It was proposed by
\citet{Kumar:2009,Kumar:2010}, in which the prompt MeV phase is
assumed to be the result of an unspecified mechanism, that the GeV LAT
emission arises from a separate exernal shock with a synchrotron peak
on the order of a few 100 MeV. Conditions on the density and the magnetic
field are then derived such that this external shock is not more
luminous than the prompt emission in the sub-MeV band. However, for
GRB~\textit{141028A} we show that the entire emission can be explained
by synchrotron emission with an MeV peak energy from a single forward
shock without fine-tuning of the parameters, i.e., both the MeV and
the delayed GeV can be explained self-consistently by an external
shock.

\subsection{Summary of Prompt Phase}
\label{sec:sumprompt}
In this section, we concisely summarize the novel
approach we have employed to test the external shock model in the
prompt phase. Almost as important as the observed $\Ep$ evolution is the
fact that we can fit the spectrum with a physical slow-cooling
synchrotron model. Even though it is possible to fit a physical
synchrotron model to the spectra (as opposed to an empirical Band
function from which the comparison to physical models must be inferred
and can lead to problems as shown in \citet{Burgess:2014b}), it must be
shown that the photon model is sound, i.e, that the dynamical evolution
can lead to slow-cooling synchrotron emission. In Section
\ref{sec:afterglow}, we confirmed that both the afterglow and prompt
phase are consistent with slow-cooling emission.

The evolution of $\Ep$ in time can be fit with the analytic prediction
of \citet{Dermer:1999}. While the blast-wave appears to be
transitioning to the asymptotic limits, ($g<1.5$) we tentatively
conclude that it is evolving towards an adiabatic regime. Full
temporal coverage of the flux and spectra through the late time
evolution will help to resolve this issue. Unfortunately, few single
pulsed GRBs have been observed with such a temporal and wavelength
coverage. This unique opportunity presents itself only due to the
current multi-mission capabilities and begs for continued
multi-wavelength coverage.

If we take the fitted parameters of the $\Ep$ evolution fit, we can predict
the evolution of the $P_{\rm p}-\Ep$ plane which is consistent with
the data in both the rise and decay phase of the pulse. The HIC of the
prompt decay phase of the synchrotron luminosity is consistent with
what would be expected from slow-cooling synchrotron in a decelerating
external shock.

We use the observation of the photospheric component to estimate both
the width of the blast-wave as well as when the relativistic reverse
shock crosses the forward shock. From our calculations, the reverse
shock will have very little impact on the evolution of the forward
shock and we therefore neglect it.

\section{Conclusion}

In this article, we have shown that the prompt phase of
GRB~\textit{141028A} is consistent with originating from an external
shock that emits synchrotron radiation from an electron distribution
that did not have enough time to cool completely
(slow-cooled). Combining the results of the $\Ep$ evolution fits with
the other clues from the data, an external shock origin of
GRB~\textit{141028A} is a very likely scenario. Not only do we find
that the prompt emission is explained by synchrotron emission from a
forward shock, but we find clues for a late time high-energy emission
resulting from the extension of this synchrotron emission rather than
an SSC component. We want to stress that in this scenario the
delineation between ``prompt'' and ``afterglow'' is superfluous as the
early and late time emission both originate from the same
mechanism. We are currently applying this analysis to other long,
single pulse GRBs and find that they too have their $\vFv$ peak and
$P_{\rm p}(t)$ which evolved consistently with the predictions of an
external shock. These results will be presented in a forthcoming
publication.

We propose that external shocks are still a viable candidate to
explain the prompt dynamics of GRBs. However, the most likely scenario
is that they are a subset of multiple dynamical frameworks including
internal shocks that further subdivide into categories based on
opacity and/or magnetic content of the outflow. We note that this is
not an entirely new idea. \citet{Panaitescu:1998} numerically
investigated the parameter space of the external shock model and
proposed that smooth, single-pulsed (and possibly multi-pulsed) GRBs
can belong to a subclass of GRBs that are the results of external
shocks. We have now shown this quantitatively by applying the model to
the data.

The analysis of individual GRBs is crucial to identifying what is
likely a host of different emission mechanisms. For example,
GRB~\textit{090902B} exhibits dynamics and spectra that are consistent
with subphotospheric dissipation \citep{Ryde:2010}. The fact that both
external shocks and photospheric emission are observationally viable
candidates for explaining GRB emission implies that we should use
caution when trying to apply one mechanism as an explanation for all
GRBs based of properties from catalogs and should instead focus on how
to relate the different emission mechanisms into a unified framework.

\acknowledgments 

The authors are very grateful for insightful discussions with Chuck
Dermer, Peter {\Mesz}, Gregory Vereshchagin and Peter Veres that
helped to improve the manuscript. This work made use of data supplied
by the UK Swift Science Data Centre at the University of Leicester,
data obtained from the Chandra Data Archive, and software provided by
the Chandra X-ray Center (CXC). JLR and AC acknowledge support for
this work from NASA through Chandra Award Number GO4-15073Z issued by
the Chandra X-ray Observatory Center under contract NAS8-03060.

The \textit{Fermi} LAT Collaboration acknowledges generous ongoing
support from a number of agencies and institutes that have supported
both the development and the operation of the LAT as well as
scientific data analysis.  These include the National Aeronautics and
Space Administration and the Department of Energy in the United
States, the Commissariat \`a l'Energie Atomique and the Centre
National de la Recherche Scientifique / Institut National de Physique
Nucl\'eaire et de Physique des Particules in France, the Agenzia
Spaziale Italiana and the Istituto Nazionale di Fisica Nucleare in
Italy, the Ministry of Education, Culture, Sports, Science and
Technology (MEXT), High Energy Accelerator Research Organization (KEK)
and Japan Aerospace Exploration Agency (JAXA) in Japan, and the
K.~A.~Wallenberg Foundation, the Swedish Research Council and the
Swedish National Space Board in Sweden.

Additional support for science analysis during the operations phase is
gratefully acknowledged from the Istituto Nazionale di Astrofisica in
Italy and the Centre National d'\'Etudes Spatiales in France.

This research made use of {\tt Astropy}, a community-developed core
Python package for Astronomy \citep{astropy} as well {\tt Matplotlib},
an open source Python graphics environment \citep{Hunter:2007}. We
also thank the developer of {\tt pymultinest} for enabling the use of
{\tt MULTINEST} in the Python environment \citep{Buchner:2014}. All
computational work was performed with the Mac OS X operating system via
the Intel Core \textit{i7} architecture.

%\bibliography{mybib}
\bibliographystyle{apj}

\appendix

\section{LAT Analysis}
\label{sec:latinfo}
The model used in the likelihood fit is composed of the Galactic diffuse
emission produced by cosmic-ray interaction with gas and radiation
fields and the isotropic diffuse emission. In addition, we add all the
point sources in the ROI with spectral models and parameters from the
2FGL catalog \citep{2012yCat..21990031N}. While the normalization of
the Galactic template and the spectral parameters of all the 2FGL
sources are frozen to their nominal values, the normalization of the
isotropic template is left free to vary in order to absorb statistical
fluctuations. The GRB location is fixed, and its spectrum is described
by a power law $dN/dE \propto E^{\gamma_{\rm ph}}$ with $\gamma_{\rm
  ph}$ the photon index (note that typically $\gamma_{\rm ph}<$0).
Following the time-resolved analysis described in
\citet{2013ApJS..209...11A} we split the LAT data in 48 log-spaced
time bins from 0.01 to 10000 seconds after the trigger.  The
logarithmically-spaced binning provides constant-fluence bins when
applied to a signal that decreases approximately as 1/time, such as
the extended GRB emissions observed by the LAT.  We first merge
consecutive time bins in order to have at least 5 counts per bin
(corresponding to the number of free parameters in the likelihood
model plus 2), and we then perform likelihood analysis.  We estimate
the significance of the GRB source by evaluating the ``Test
Statistic'' (TS) equal to twice the logarithm of the ratio of the
maximum likelihood value produced with a model including the GRB over
the maximum likelihood value of the null hypothesis, i.e., a model
that does not include the GRB. The probability distribution function
(PDF) of the TS under the null hypothesis is given by the probability that
a measured signal is compatible with statistical fluctuations.  The
PDF in such a source-over-background model cannot, in general, be
described by the usual asymptotic distributions expected from Wilks'
theorem~\citep{Wilks:38,Protassov02}.  However, it has been verified
by dedicated Monte Carlo simulations~\citep{Mattox96} that the
cumulative PDF of the TS in the null hypothesis (i.e., integral of the
TS PDF from some TS value to infinity) is approximately equal to a
$\chi^2_{n_{\rm dof}}/2$ distribution, where $n_{\rm dof}$ is the
number of degrees of freedom associated with the GRB.  The factor of
1/2 in front of the TS PDF formula results from allowing only positive
source fluxes.

If the resulting TS value is lower than an arbitrary threshold
($TS<10$) we merge the corresponding time bin with the next one, and
we repeat the likelihood analysis. This step is iterated until one of
two conditions is satisfied: 1) we reach the end of a GTI before
reaching $TS = 10$, in which case we compute the value of the 95\% CL
upper limit (UL) for the flux evaluated using a photon index of $-$2;
2) we reach $TS > 10$, in which case we evaluate the best-fit values
of the flux and the spectral index along with their 1$\sigma$ errors.

\section{High-Energy  SSC Fraction }
\label{sec:sscCalc}

Following \citet{Beniamini:2013}, the SSC and Klein-Nishina peak
frequencies are defined respectively as

\begin{align}
  \label{eq:pf}
&\nu_{\rm min} = \Ep h^{-1}\\
&\nu_{\SSC} = \gamma_{\rm min}^2 \nu_{\rm min}\\
&\nu_{\KN}  = \Gamma \gamma_{\rm min} m_{\rm e} c^2 h^{-1}
\end{align}

\noindent where $h$ is Planck's constant. The total SSC flux
is linked to the total synchrotron flux at the peak ($F_{\nu_{\rm
    min}}^{\synch}$) by

\begin{equation}
  \label{eq:link}
  \frac{F_{\SSC}}{\nu_{\rm min}F_{\nu_{\rm min}}^{\synch} } = Y \Lambda_{\KN}
\end{equation}

\noindent where $Y$ is the Compton parameter and,

\begin{equation}
  \label{eq:lkn}
  \Lambda_{\KN}  =\left\{
\begin{array}{ll}
 \left(\frac{\nu_{\SSC}}{\nu_{\KN}}\right)^{-\nicefrac{1}{2}} & \frac{\nu_{\SSC}}{\nu_{\KN}}>1 \\
 	1                                                &  \text{otherwise}%
\end{array}
\right.\text{.}
\end{equation}

\noindent In the derivation of the $Y$ via \citet{Sari:1996}, the
Klein-Nishina correction of the Compton cross-section ($\sigma_{\KN}
\propto \ln(2 x)/x$ where $x$ is the photon energy in the rest frame
of the electron in $m_{\rm e} c^2$ units) was ignored. Here, in order
to estimate $Y$ we include this correction by considering electrons
with Lorentz factor $\gamma_{\rm min}$ and their corresponding
synchrotron photons. Therefore, we write
\begin{equation}
  \label{eq:compparam}
  Y= \left( \frac{\epsilon_{\rm e}}{\epsilon_{\rm B}}\right)^{\nicefrac{1}{2}}\cdot  \left[\frac{\ln \left( 2 \frac{\gamma_{\rm min} h \nu_{\rm min}}{m_{\rm e} c^2} \right)}{\gamma_{\rm min}   \frac{ h \nu_{\rm min}}{m_{\rm e} c^2}}\right]^{\nicefrac{1}{2}}\;\text{.}
\end{equation}

\noindent The total upscattered flux in the LAT bandpass can then be
written as

\begin{equation}
  \label{eq:totalflux}
\frac{{F_{\SSC}^{\LAT}}}{\nu_{\rm min}F_{\nu_{\rm min}}^{\synch}} = \frac{{F_{\SSC}}}{\nu_{\rm min}F_{\nu_{\rm min}}^{\synch}}\cdot \Lambda_{\rm W}
\end{equation}

\noindent where

\begin{equation}
  \label{eq:w}
 \Lambda_{\rm W}  =\left\{
\begin{array}{ll}
                  1                                         & \text{min}(\nu_{\SSC}\text{,}\nu_{\KN})< \nu_{\rm max} \\
 \left(\frac{\text{min}(\nu_{\SSC}\text{,}\nu_{\KN})}{\nu_{\rm max}}\right)^{-\alpha-2} & \text{otherwise} 
\end{array}
\right. \text{.}
\end{equation}

\noindent Here, $\alpha$ refers to the low-energy photon index of the
Band function which is -2/3 in our case and $\nu_{\rm max}$ is the
maximum synchrotron frequency corresponding to the maximum Lorentz
factor of the accelerated electrons \citep{Jager:1996}:

\begin{equation}
  \label{eq:gmax}
\gamma_{\rm max} = 4\cdot10^{7}\left(\frac{B}{1 \; \text{G}} \right)^{-\nicefrac{1}{2}}
\end{equation}

\noindent In order to estimate the synchrotron flux in the LAT band
pass, we write

\begin{equation}
  \label{eq:synflux}
  F_{\nu}^{\synch} = \left(\frac{\nu}{\nu_{\rm min}} \right)^{-\frac{s-1}{2}} F_{\nu_{\rm min}}^{\synch}
\end{equation}

\noindent where $s$ is the high-energy electron power law index which
we take $s\equiv 2.5$ for simplicity. Then we can write the fraction
of SSC to synchrotron flux in the LAT bandpass as

\begin{equation}
  \label{eq:fracit}
  \mathcal{F}_{\SSC}\equiv \frac{F_{\SSC}^{\LAT}}{F_{\synch}^{\LAT}} = \frac{F_{\SSC}^{\LAT}}{\int_{\LAT} d\nu\;\left(\frac{\nu}{\nu_{\rm min}} \right)^{-\frac{s-1}{2}}F_{\nu_{\rm min}}^{\synch}} 
\end{equation}

\noindent This can be further simplified via Equations \ref{eq:link}
and \ref{eq:totalflux} to

\begin{equation}
  \label{eq:fracit2}
  \mathcal{F}_{\SSC} = \frac{\nu_{\rm min}}{\int_{\LAT} d\nu\;\left(\frac{\nu}{\nu_{\rm min}} \right)^{-\frac{s-1}{2}}}\cdot \Lambda_{\rm W} \cdot \Lambda_{\KN} \cdot  Y \text{.}
\end{equation}

\noindent This is in fact an upper limit on the fraction of SSC due to
the assumption that electrons radiate all their energy in a dynamical
time either by synchrotron \textit{and/or inverse-Compton} and that
there is a significant fraction of electrons below $\gamma_{\rm
  min}$. If either assumption fails (as is true in our analysis) the
fraction of SSC emission will be suppressed.

\clearpage

\end{document}